\documentclass[superscriptaddress,groupedaddress,nofootnoteinbib,11pt]{article}

\usepackage{jcappub}
\usepackage{bm}
\usepackage{verbatim}

\usepackage{epsf}
\usepackage{graphicx,epsfig}
\usepackage{amsfonts}
\usepackage{amssymb}
\usepackage{xcolor}

\definecolor{mine}{rgb}{0.2,0.1,0.7}
\definecolor{bb}{rgb}{0.3, 0.5, 1}
\definecolor{bg}{rgb}{0.1, 0.1, 0.5}

\def\M{M_{\rm Pl}}

\def\La{\Lambda_3}

\def\L{\Lambda}
\def\A{A}

\def\tp{\tilde{\pi}}

\def\L*{{\cal L}_*}
\def\L{\mathcal{L}}
\def\({\left(}
\def\){\right)}

\def\nn{\nonumber}
\def\mn{_{\mu \nu}}
\def\stu{St\"uckelberg }

\def\<{\langle}
\def\>{\rangle}
\def\Ein{\hat{\mathcal{E}}}

\newcommand{\bea}{\begin{eqnarray}}
\newcommand{\eea}{\end{eqnarray}}
\newcommand\be{\begin{equation}}
\newcommand\ee{\end{equation}}
\newcommand\beq{\begin{equation}}
\newcommand\eeq{\end{equation}}

\def\ba{\begin{eqnarray}}
\def\ea{\end{eqnarray}}

\def\n{\bar{\nu}}
\def\l{\bar{\lambda}}

\newcommand{\refeq}[1]{(\ref{#1})}

\begin{document}

\title{On the Vainshtein mechanism in the minimal model of massive gravity}

\author[1,2]{S\'ebastien Renaux-Petel}
\affiliation[1]{Laboratoire de Physique Th\'eorique et Hautes
   Energies, Universit\'e  Pierre \& Marie Curie - Paris VI, CNRS-UMR 7589, 4 place Jussieu, 75252 Paris, France}
\affiliation[2]{Sorbonne Universit\'es, Institut Lagrange de Paris,
  98 bis Bd Arago, 75014 Paris, France}

\vskip 4pt

\date{\today}


\abstract{We reinvestigate the fate of the Vainhstein mechanism in the minimal model of dRGT massive gravity. As the latter is characterised by the complete absence of interactions in the decoupling limit, we study their structure at higher energies. We show that in static spherically symmetric configurations, the lowest energy scale of interactions is pushed up to the Planck mass. This fact points towards an absence of Vainshtein mechanism in this framework, but does not prove it. By resorting to the exact vacuum equations of motion, we show that there is indeed an obstruction that precludes any recovery of General Relativity under the conditions of stationarity and spherical symmetry. However, we argue that the latter are too restrictive and might miss some important physical phenomena. Indeed, we point out
that in generic non spherically symmetric or time-dependent situations, interactions arising at energies arbitrarily close to the energy scale of the decoupling limit reappear. This leads us to question whether the small degree of spherical symmetry breaking in the solar system can be sufficient to give rise to a successful Vainshtein mechanism.}

\maketitle

\section{Introduction}

How to give a mass to the graviton is an interesting theoretical question in its own right. Moreover, achieving it in a controlled manner might help to tackle the cosmological constant problem \cite{Weinberg:1988cp}, by weakening gravity on cosmological scales through the degravitation mechanism \cite{Dvali:2002pe,ArkaniHamed:2002fu,Dvali:2007kt}. However, such a modification usually comes with pathologies, in particular the presence of the so-called Boulware-Deser ghost \cite{Boulware:1973my,Creminelli:2005qk,Deffayet:2005ys}. It is only recently that this problem has been solved by de Rham, Gabadadze and Tolley \cite{deRham:2010ik,deRham:2010kj}, who formulated a ghost-free non-linear theory of massive gravity, henceforth dRGT. The absence of ghosts in this theory has now been confirmed and formulated by several authors in several formalisms, see \textit{e.g.} Refs.~\cite{Hassan:2011hr,deRham:2011rn,deRham:2011qq,Hassan:2011ea,Mirbabayi:2011aa,Golovnev:2011aa,Hassan:2012qv,Kluson:2012wf,Deffayet:2012zc}. However, for a theory of massive gravity to be viable, not only should it be devoid of ghosts, but it should also conform with gravity precision tests in the solar system, where no deviation from General Relativity (GR) is detected \cite{Will:2005va}. The screening of the additional degrees of freedom of massive gravity compared to GR near dense sources has proved to be a non-trivial task. It is inefficient at the linear level, a manifestation of the so-called vDVZ (van Dam-Veltman-Zakharov) discontinuity \cite{vDVZ}. However, Vainshtein suggested that the non-linearities of massive gravity can yield this effect \cite{Vainshtein:1972sx}, by rendering the new degrees of freedom strongly kinetically self-coupled, so that they almost do not propagate. That this mechanism can work has been proven explicitly only recently \cite{Babichev:2009us,Babichev:2009jt,Babichev:2010jd} (see Ref.~\cite{Babichev:2013usa} for a review about the Vainshtein mechanism).

In this respect, there exist several studies of spherically symmetric solutions in dRGT massive gravity \cite{Koyama:2011xz,Nieuwenhuizen:2011sq,Koyama:2011yg,Chkareuli:2011te,Gruzinov:2011mm,Comelli:2011wq,Berezhiani:2011mt,Sjors:2011iv,Sbisa:2012zk,Volkov:2013roa}, both exactly and in the so-called decoupling limit \cite{ArkaniHamed:2002sp}. The latter is defined in such a way as to concentrate on the interactions arising at the lowest energy scale, which, in generic dRGT theories, is identified as $\La=(\M m^2)^{1/3}$, where $m$ is the mass of the graviton. In this paper, we reinvestigate the fate of the Vainshtein mechanism in the so-called minimal model of massive gravity, a particular model defined by the property that its decoupling limit Lagrangian is trivial, \textit{i.e.} that no interactions arise at the energy $\La$. In Ref.~\cite{Koyama:2011yg}, it is argued that the Vainshtein mechanism is ineffective in this model, and that the latter is therefore ruled out by solar system observations. However, the arguments there are based on a weak field approximation for the helicity-2 mode, which ultimately captures the picture of the decoupling limit. Yet, the absence of non-linearities at this level does not necessarily mean that they are ineffective to yield a Vainshtein mechanism. It solely implies that the decoupling limit Lagrangian does not suffice
 to decipher its existence or not in this model. Additionally, the analysis of Ref.~\cite{Koyama:2011yg} is restricted to spherically symmetric configurations. While this is a natural starting point and a very common assumption in the literature, we will see that the peculiarities of the minimal model might render it misleading as to the fate of the Vainshtein mechanism in real-world conditions, in which spherical symmetry is broken, even mildly, like in the solar system.

For these reasons, we investigate the interactions of the minimal model arising at energies higher than $\La$, in static spherically symmetric configurations and beyond. Working within the \stu formalism, and considering only the scalar graviton, we prove
 the remarkable fact that in static spherically symmetric configurations, all interactions arising below the Planck mass vanish identically, \textit{i.e.} beyond the free, quadratic, action, the theory completely loses track of the graviton mass, barring energies larger than the Planck scale. This tantalizing fact points towards an absence of Vainshtein mechanism in this set up, but does not prove it. For this reason, we resort to the exact equations of motion in the metric formalism. We then show completely generally that in all vacuum stationary and spherically symmetric configurations, there exists an obstruction that precludes any recovery of General Relativity. While this could be seen as a proof that the minimal model is ruled out, we argue that our analysis of the energy scales of interactions in time-dependent or non-spherically symmetric configurations indicates that it would be premature to reach this conclusion without further study.\\

The outline of the paper is as follows. In section \ref{Minimal}, we introduce the theory of dRGT massive gravity, review the construction of its decoupling limit, and present the minimal model. In section \ref{Energy}, we investigate, within the \stu formalism, energy scales of interactions in the minimal model beyond the decoupling limit. We study exact vacuum solutions of the theory in section \ref{Exact} and conclude in section \ref{Conclusion}.

\section{dRGT massive gravity and its minimal model}
\label{Minimal}

In this paper, we consider ghost-free dRGT massive gravity with Minkowski reference metric in four spacetime dimensions. Its Lagrangian is usually formulated as \cite{deRham:2010kj}
\be
{\cal L}_{{\rm MG}}=\sqrt{-g} \frac{\M^2}{2} \left(R +m^2 \left( \L^{(2)}({\cal K}) +\alpha_3  \L^{(3)}({\cal K})+\alpha_4 \L^{(4)}({\cal K})  \right) \right)
\label{L-alpha}
\ee
where $R$ is the Ricci scalar of the spacetime metric $g_{\mu \nu}$, $m$ is the mass of the graviton, and
\ba
\label{U2}
\L^{(2)}({\cal K})&=&  \langle \mathcal{K}\rangle^2- \langle \mathcal{K}^2\rangle,\\
\label{U3}
\L^{(3)}({\cal K})&=&  \langle \mathcal{K}\rangle^3-3 \langle \mathcal{K}\rangle  \langle\mathcal{K}^2\rangle+2 \langle\mathcal{K}^3\rangle,\\
\label{U4}
\L^{(4)}({\cal K})&=&  \langle\mathcal{K}\rangle^4-6 \langle\mathcal{K}^2\rangle \langle\mathcal{K}\rangle^2+8 \langle\mathcal{K}^3\rangle \langle\mathcal{K}\rangle+3 \langle\mathcal{K}^2\rangle^2-6 \langle\mathcal{K}^4\rangle\,,
\ea
where $\langle \ldots \rangle$ represents the trace of a tensor and ${\cal K}^{\mu}{}_{\nu}=\delta^{\mu}{}_{\nu}-\gamma^{\mu}{}_{\nu}$, with
\be
\gamma^{\mu}{}_{\nu} =\sqrt{g^{\mu \alpha} \eta_{\alpha \nu}}\,.
\ee
Here, $g^{\mu \nu}$ is the inverse of $g_{\mu \nu}$ and the square root is understood in the matrix sense, \textit{i.e.} 
\be
(\gamma^2)^{\mu}{}_{\nu} \equiv \gamma^{\mu}{}_{\alpha}   \gamma^{\alpha}{}_{\nu} =g^{\mu \alpha} \eta_{\alpha \nu}\,.
\ee
We will only be concerned with cases in which $g_{\mu \nu}$ is close to $\eta_{\mu \nu}$, so that $g^{\mu \alpha} \eta_{\alpha \nu}$ is close to the identity matrix, and the matrix square root is well defined by perturbation theory (see \cite{Deffayet:2012zc,Gratia:2013gka} for discussions of more general cases). The formulation \refeq{L-alpha} is useful because it renders explicit the fact that, upon expanding $g_{\mu \nu}$ about the Minkowski metric, the Fierz-Pauli structure \cite{Fierz:1939ix} is recovered at the level of the quadratic action.

The mass term in Eq.~\refeq{L-alpha} explicitly breaks the covariance of General Relativity. However, it can be usefully restored by resorting to the well-known \stu trick (see \textit{e.g.} \cite{Siegel:1993sk,ArkaniHamed:2002sp,Creminelli:2005qk} for more details).  In practice, it amounts to promoting the fixed metric $\eta\mn$ to a tensor field, through the replacements:
\bea
 \eta\mn  &\to& \tilde \eta\mn= \eta\mn -\nabla_{\mu} V_{\nu} -\nabla_{\nu} V_{\mu} +\eta^{\alpha \beta} \nabla_{\mu} V_{\alpha} \nabla_{\nu} V_{\beta}\,, \label{stu1} \\
V_{\mu} &\to& \tilde A_{\mu} +\nabla_{\mu} \tilde \pi\,, \label{stu2}
\eea
where $\nabla_{\mu}$ denotes the covariant derivative with respect to the reference Minkowski metric expressed in any coordinate system (it will be useful later on when dealing with spherically symmetric configurations). Upon performing these replacements, the counting of degrees of freedom (d.o.f) becomes more transparent. Indeed, instead of having all d.o.f of massive gravity contained in the dynamical metric $g\mn$, its fluctuations about the Minkowski background, $\tilde h\mn \equiv g\mn-\eta\mn$, now encode only the $2$ d.o.f of standard General Relativity (up to a field redefinition, see below), whereas $\tilde A_{\mu}$ is a vector bearing 2 d.o.f and $\tilde \pi$ is a scalar (often called the scalar graviton) which, in dRGT theories, obeys second-order equations of motion and hence contains one d.of. In terms of these variables, a linear analysis (dictated solely but the Fierz-Pauli structure) then reveals that the canonically normalised fields read
\be
h\mn = \M \tilde h\mn\,, \quad A_{\mu} =\M m \tilde A_{\mu}\,, \quad \pi=\M m^2 \tilde \pi\,.
\ee
As for interactions, the structure of Eqs.~\refeq{stu1}-\refeq{stu2} is such that in the mass term, $\pi$ always appears with two (covariant) derivatives, $\A$ with one, and $h$ with none, so that a generic interacting term schematically reads
\be
 \sim m^2 \M^2 \tilde h^{n_h} (\partial \tilde A)^{n_A} (\partial^2 \tilde \pi)^{n_{\pi}} \sim \Lambda_{\alpha}^{4-n_h-2n_A-3n_{\pi}} h^{n_h} (\partial A)^{n_A} (\partial^2 \pi)^{n_{\pi}}\,,
 \label{interactions}
\ee
where the energy scale suppressing each term is
\be
\Lambda_{\alpha}=\left(\M m^{\alpha-1} \right)^{1/\alpha}\,, \qquad \alpha=\frac{3n_{\pi}+2n_A+n_h-4}{n_{\pi}+n_{A}+n_h-2}\,,
\label{Lambda-lambda}
\ee
and where $n_{\pi}+n_A+n_h \geq 3$ since we are considering interactions. Since $m < \M$ for realistic parameters, $\Lambda_{\alpha}$ is a decreasing function of $\alpha$. For a generic mass term, there are interactions whose energy scales lie below $\Lambda_3$. However, the dRGT Lagrangian \refeq{L-alpha} is built such that the corresponding terms, in $(\partial^2 \pi)^{n_{\pi}}$ and in $(\partial A) (\partial^2 \pi)^{n_{\pi}}$, vanish identically. The lowest interaction scale is thus in general $\La=(\M m^2)^{1/3}$. By considering the so called \textit{decoupling limit} (DL), such that
\be
m \to 0\,, \quad \M \to \infty\,, \quad \La\,\, {\rm fixed}\,,
\label{DL}
\ee
one can therefore concentrate on the leading interactions of the theory. In this limit, the non-linearities in $h\mn$ and $A_{\mu}$ disappear, and one can show that the full Lagrangian \refeq{L-alpha} boils down to (ignoring the free field $A_\mu$ in this limit) \cite{deRham:2010ik,deRham:2010kj}
\ba
\label{DL1}
\L_{{\rm DL}}&=& -\frac14 h^{\mu \nu} \Ein^{\alpha \beta}\mn h_{\alpha \beta}
+\frac12 h^{\mu \nu}  \left( 2 X^{(1)}\mn  +(1+3 \alpha_3)\frac{X^{(2)}\mn }{ \La^{3}}+(\alpha_3+4 \alpha_4)\frac{X^{(3)}\mn }{ \La^{6}} \right)\,,
\ea
where $\Ein$ is the Lichnerowicz operator, coming from the expansion of the Einstein-Hilbert action at quadratic order, and
 \ba
 \label{Xmn}
 X^{(n)}\mn=\sum_{m=0}^n (-1)^m \frac{n!}{2(n-m)!} (\Pi^m)\mn \L^{(n-m)}(\Pi)\,,
 \ea
 where $\Pi\mn \equiv \nabla_{\mu} \nabla_{\nu} \pi$, $\L^{(0)}\equiv1$, $\L^{(1)}(\Pi)=\langle \Pi \rangle$ and $\L^{(2,3,4)}$ are defined in Eqs.~\refeq{U2}-\refeq{U4}. 
 
 The Lagrangian \refeq{DL1} kinetically mixes $h\mn$ and $\pi$. One can partially diagonalize it by use of the transformation (the mixing in $X^{(3)}\mn$ can be eliminated as well but at the cost of a non-local field redefinition \cite{deRham:2010ik})
\ba
h\mn=\bar h\mn + \pi\, \eta\mn -\frac{1+3 \alpha_3}{\La^3}\nabla_\mu \pi \nabla_\nu \pi\,,
\label{redefinition}
\ea
yielding
\ba
\L_{{\rm DL}}&=& -\frac14 \bar h^{\mu \nu} \Ein^{\alpha \beta}\mn \bar h_{\alpha \beta}
+\frac12  (\alpha_3+4 \alpha_4) \bar h^{\mu \nu} \frac{X^{(3)}\mn }{ \La^{6}}  +\sum_{n=2}^{5} c_n  \frac{\L^{(n)}_{\rm Gal}  }{ \La^{3(n-2)} } \,,
\label{final}
\label{DL2}
\ea
where
\ba
c_n&=& -\frac34\delta_{n,2} -\frac34(1+3 \alpha_3) \delta_{n,3}- \left(\frac14 (1+3 \alpha_3)^2+\frac12(\alpha_3+4 \alpha_4) \right) \delta_{n,4} \nonumber \\
&&-\,\,  \frac{5}{8}  (1+3 \alpha_3)(\alpha_{3}+4\alpha_{4}) \delta_{n,5}
\ea
and where the Galileon Lagrangians are such that $\L^{(n)}_{\rm Gal} \equiv (\nabla \pi)^2 \L^{(n-2)}(\Pi)$.\\

From the above formulas, it is clear that the so-called minimal model, such that $1+3\alpha_3=\alpha_3+4 \alpha_4=0$ is very peculiar\footnote{Within the context of Galileons, or equivalently at the level of the DL action \refeq{DL1}, it has been shown in Ref.~\cite{Berezhiani:2013dw} that no Vainshtein mechanism is possible unless $1+3 \alpha_3>0$, and unless $\alpha_3+4\alpha_4=0$ in Ref.~\cite{Berezhiani:2013dca}. However, the case of the minimal model, for which both these parameters vanish and one should go beyond the DL, was not considered in these papers. We thank Claudia de Rham for pointing them out to us.}. Indeed, for this particular choice of parameters, all interactions in the decoupling limit Lagrangian vanish identically, \textit{i.e.} $\bar h\mn$ and $\pi$ are just free fields in this limit. The appearance of these particular combination of the parameters $\alpha_3$ and $\alpha_4$ can be understood non-perturbatively by formulating the Lagrangian \refeq{L-alpha} in terms of the $\L^{(n)}(\gamma)$ instead of the $\L^{(n)}(\cal K)$, where we recall that ${\cal K}^{\mu}{}_{\nu}=\delta^{\mu}{}_{\nu}-\gamma^{\mu}{}_{\nu}$. One then finds
\be
{\cal L}_{{\rm MG}}=\sqrt{-g} \frac{\M^2}{2} \left(R +m^2 \sum_{n=0}^4 \beta_n\, \L^{(n)}(\gamma) \right)
\label{L-beta}
\ee
where
\bea
\beta_0&=&12 \left(1+2(\alpha_3+\alpha_4) \right) \\
\beta_1&=&-6 \left(1+3 \alpha_3+4 \alpha_4 \right) \\
\beta_2&=&1+3 \alpha_3+3(\alpha_3+4 \alpha_4)  \label{beta2} \\
\beta_3&=&-\left(\alpha_3+4 \alpha_4 \right) \label{beta3} \\
\beta_4&=&\alpha_4\,.
\eea
The term in $\beta_4$, proportional to $\sqrt{-g} \L^{(4)}(\gamma) =24 \sqrt{-g} \,{\rm det}(g^{-1} \eta)=24 \sqrt{-{\rm det }(\eta_{\mu \nu}) }$, is non-dynamical and can be omitted. The full Lagrangian of the minimal model, such that $1+3\alpha_3=\alpha_3+4 \alpha_4=0$, can thus be rewritten in the form
\bea
\L_{{\rm min}}&=&\sqrt{-g} \frac{\M^2}{2} \left(R +2 m^2 \left(3 - \langle \gamma \rangle \right) \right) \label{L-min} \\
&\equiv &\sqrt{-g} \frac{\M^2}{2} R + {\cal L}_{{\rm mass}} \label{L-mass}\,.
\eea
Contrary to the formulation \refeq{L-alpha}, the Fierz-Pauli structure at the level of the quadratic action is not obvious in this language. However, we will see in what follows that the absence of terms quadratic or higher-order in $\gamma$ in ${\cal L}_{{\rm mass}}$ considerably simplifies the discussion of the interactions in this model, as well as it plays a crucial role in the obstruction to obtain a Vainshtein mechanism in stationary and spherically symmetric configurations.

\section{Energy scales of interactions in the minimal model}
\label{Energy}

As we have seen in the previous section, the decoupling limit Lagrangian of the minimal model of dRGT massive gravity is trivial, \textit{i.e.} $\bar h\mn$ and $\pi$ are just free fields in this limit. This merely shows that the decoupling limit \refeq{DL} is not adequate to describe interactions in the minimal model, and that those arising at energy scales higher than $\Lambda_3$ should be determined and taken into account. For this reason, we explore in this section the structure of the interactions between $\bar h\mn$ and $\pi$. From Eqs.~\refeq{stu1}-\refeq{stu2}-\refeq{redefinition}, we thus write
\bea
g\mn&=&\eta\mn+\frac{\bar h\mn}{\M}+\frac{\pi}{\M} \eta\mn \,, \\
\tilde \eta\mn &=& \eta\mn -\frac{2}{\La^3} \Pi\mn + \frac{1}{\La^6} \Pi_{\mu \alpha} \eta^{\alpha \beta}  \Pi_{\beta \nu} \,. \label{eta-tilde}
\eea
As our primary interest is in static spherically symmetric (SSS) configurations, we choose the Schwarzschild gauge for $\bar h\mn$, writing
\be
g_{\mu \nu} dx^{\mu} dx^{\nu} = -\left(1+\frac{\n}{\M}\right) dt^2+\left( 1+\frac{\l}{\M} \right) dr^2+r^2 d \Omega^2+\frac{\pi}{\M} \left(-dt^2+dr^2+r^2 d\Omega^2 \right)\,,
\label{g} 
\ee
where $d \Omega^2=d \theta^2+{\rm sin}^2 \theta \, d \phi^2$ and where $\n, \l$ and $\pi$ are functions of $r$ only. From Eq.~\refeq{eta-tilde} we obtain 
\be
\tilde \eta_{\mu \nu} dx^{\mu} dx^{\nu} = - dt^2+ \left(1-\frac{\pi''}{\La^3} \right)^2  dr^2+ \left(1-\frac{\pi'}{r \La^3}\right)^2 r^2 d \Omega^2 
\label{eta-square}
\ee
where $' \equiv d/dr$. An obvious matrix square root $\gamma^{\mu}{}_{\nu}$ of $g^{\mu \alpha} \tilde \eta_{\alpha \nu}$ thus reads, in the coordinate system $(t,r,\theta,\phi)$:
\bea
\gamma^{\mu}{}_{\nu}&=& {\rm Diag} \left( \left(-g_{tt} \right)^{-1/2} , \, \left(g_{rr}\right)^{-1/2} \left(1-\frac{\pi''}{\La^3} \right) ,  \right.
\nn
\\
&& \left. \left( 1+\frac{\pi}{\M}\right)^{-1/2}  \left(1-\frac{\pi'}{r \La^3}\right) , \, \left( 1+\frac{\pi}{\M}\right)^{-1/2}  \left(1-\frac{\pi'}{r \La^3}\right) \right)
\label{gamma}
\eea 
with 
\bea
-g_{tt}=1+\frac{\n+\pi}{\M} \qquad {\rm and}\qquad  g_{rr}=1+\frac{\l+\pi}{\M}\,.
\eea
The remarkable point here is that in SSS configurations, the matrix structure in Eq.~\refeq{eta-tilde} trivializes to lead to the perfect-square structure \refeq{eta-square}. As a consequence, the matrix $\gamma$ in Eq.~\refeq{gamma} contains only one power of $\pi/\La^3$ (and its derivatives), in contrast with the generic situation where taking the square root of $g^{\mu \alpha} \tilde \eta_{\alpha \nu}$ generates an infinite number of them (see below).\\

With the explicit expression \refeq{gamma}, one obtains the expression of the mass term \refeq{L-mass}:
\bea
{\cal L}_{{\rm mass}}
&=& \M \La^3 r^2 \left(1+\frac{\pi}{\M} \right) \left( -g_{tt} g_{rr} \right)^{1/2}   \nn \\
&\times& \left[ 3-(-g_{tt})^{-1/2}-g_{rr}^{-1/2}  \left(1-\frac{\pi''}{\La^3} \right)-2  \left(1+\frac{\pi}{\M} \right) ^{-1/2}  \left(1-\frac{\pi'}{r \La^3}\right)  \right]\,.
\label{explicit-mass-term}
\eea
The factors of $1/\La^3$ in $\pi''/\La^3$ and $\pi'/(r \La^3)$ are compensated by the overall factor $\M^2 m^2=\M \La^3$, whereas only $\M$ enters into the factors $1+\pi/\M, g_{tt}$ and $g_{rr}$. As a consequence, the interactions in \refeq{explicit-mass-term} are either suppressed by the Planck mass, or of the type $\sim \M \La^3 X^n/\M^n$, where $n \geq 3$ and $X^n$ stands for $n$ products of either $\l,\n$ or $\pi$. The energy scales suppressing the latter terms are of the type $\Lambda_{\alpha}$ \refeq{Lambda-lambda} with $\alpha=1-2/(n-2)$ and are thus above the Planck mass. As the interactions coming from the Einstein-Hilbert action are suppressed by the Planck mass (see Eq.~\refeq{g}), we reach the conclusion that the lowest energy scale of interactions is $\M$. In other words, not only do the interactions at the energy $\La$ vanish, but also all the interactions involving $\bar h\mn$ and $\pi$ below the Planck mass! Had we used the formulation \refeq{L-alpha}, this would have been obscured: each term $\L^{(n)}({\cal K})$ contains non-trivial interactions below $\M$, conspiring to cancel for the specific parameters of the minimal model $\alpha_3=-1/3, \alpha_4=1/12$. As we have seen, this becomes transparent with the formulation \refeq{L-beta}, relying on the fact that $\gamma$ is linear in $\pi/\La^3$ (this is a consequence of considering a SSS configuration) and that the minimal model \refeq{L-min} is linear in $\gamma$ itself.\\

\noindent  {\bf Beyond static spherically symmetric configurations}. One can wonder to which extent the above conclusion relies on considering a SSS configuration. To understand this, we first consider a spherically symmetric configuration of the form \refeq{g}, but now with an additional time-dependence of $\n, \l$ and $\pi$. One then obtains
\begin{displaymath}
(g^{-1} \tilde \eta)^{\mu}{}_{\nu }=\left(\begin{array}{c|c}
{\bf \A} \,& {\bf 0}  \\
\hline
{\bf 0} & \,{\bf B} \end{array}\right) 
\end{displaymath}
where 
\begin{displaymath}
{\bf \A}=\left(\begin{array}{c|c}
-g_{tt}^{-1}\left( (1+\ddot{\tp})^2-\dot{\tp}'^2 \right)& -g_{tt}^{-1}  \dot{\tp}' \left(2-\tp''+\ddot{\tp} \right)  \\
\hline
g_{rr}^{-1} \dot{\tp}' \left(2-\tp''+\ddot{\tp} \right) & g_{rr}^{-1} \left( (1-\tp'')^2-\dot{\tp}'^2 \right)\end{array}\right) \,,
\end{displaymath}
\be
{\bf B}=\left(1+\frac{\pi}{\M}\right)^{-1} \left(1-\frac{\tp'}{r}\right)^2  \bf{1}_2\,,
\ee
we recall that $\pi=\La^3 \tilde \pi$, and $\dot{} \equiv d/dt$. When $\dot{\tp}'=0$, a matrix square-root of ${\bf \A}$ is 
\be
{\rm Diag} \left( (-g_{tt})^{-1/2} (1+\ddot{\tp}) , (g_{rr})^{-1/2} (1-\tp'')\right),
\ee
then $\gamma^{\mu}{}_{\nu}$ is linear in $\tp=\pi/\La^3$ and the above arguments apply\footnote{Note in particular that the latter apply when $\pi$ is of the form $\pi=\pi_0(r)+ c \,t^2$, where $c$ is a constant, as considered for instance in Refs.~\cite{Berezhiani:2013dw,Berezhiani:2013dca}.}. Similarly, when $\n=\l$, one can find an ``obvious'' square-root of ${\bf \A}$:
\begin{displaymath}
g_{rr}^{-1/2}\left(\begin{array}{c|c}
1+\ddot{\tp} & \dot{\tp}'  \\
\hline
-\dot{\tp}' & 1-\tp'' \end{array}\right) \,,
\end{displaymath}
 $\gamma^{\mu}{}_{\nu}$ is still linear in $\pi/\La^3$ and the minimal energy of interactions is $\M$. However, this does not hold in full generality. Using that for any $2 \times 2$ matrix ${\bf \A}$, a square root {\bf R} of ${\bf \A}$ (the one connected to the identity in perturbation theory) reads
 \be
 {\bf R}=\left( \langle {\bf \A} \rangle +2 \sqrt{{\rm det}({\bf \A})} \right)^{-1/2} \left( {\bf \A}+\sqrt{{\rm det}({\bf \A})}  \bf{1}_2     \right)\,,
\ee
with its trace given by
 \be
 \langle {\bf R} \rangle=\left( \langle {\bf \A} \rangle +2 \sqrt{{\rm det}({\bf \A})} \right)^{1/2}  \,,
 \label{trace-square-root}
\ee
one arrives at the explicit expression for $\langle \gamma \rangle$:
\bea
\langle \gamma \rangle&=& \left[ g_{rr}^{-1} \left((1-\tp'')^2-\dot{\tp}'^2 \right)+(-g_{tt})^{-1} \left(  (1+\ddot{\tp})^2-\dot{\tp}'^2  \right)  \right.
\nn
\\
&& \left.
+2 \left(1+\dot{\tp}'^2+\ddot{\tp}-\tp''(1+\ddot{\tp}) \right) (- g_{tt} g_{rr})^{-1/2} \right]^{1/2}
+2 \left(1+\frac{\pi}{\M}\right)^{-1/2} \left(1-\frac{\tp'}{r}\right).
\label{gamma-t}
 \eea
Beyond interactions similar to the SSS case, suppressed by $\M$ and $\Lambda_{1-2/(n-2)}$, it is clear from the above remarks that all the other possible interactions should vanish for $\dot{\tp}'=0$ and $\n=\l$. Indeed, by using Eq.~\refeq{gamma-t} and by expanding the term $\M^2 m^2 \sqrt{-g} \langle \gamma \rangle$ in Eq.~\refeq{L-mass} in terms of the fields $\n,\l$ and $\pi$, one finds an infinite number of interactions involving negative powers of $\La^3$, all of them being proportional to $(\l-\n)^2 \dot{\tp}'^2$. Amongst them, we find for example interactions of the type
\be
a_n \frac{r^2 (\l-\n)^2 \dot{\pi}'^2}{\La^3 \M}  \frac{ \left(\pi''-\ddot{\pi} \right)^n }{\La^{3 n}}  \,,
\ee
with non-zero $a_n$ for all $n \geq 0$. Note that the structure of these terms is in agreement with the general analysis Eqs.~\refeq{interactions}-\refeq{Lambda-lambda}, with $n_h=2$ and $n_{\pi}=n+2$. In particular, the energy scales suppressing these interactions are 
\be
\Lambda_{(3n+4)/(n+2)}=\left(\M m^{\frac{2(n+1)}{n+2}} \right)^{\frac{n+2}{3n+4}}\,,
\label{E-n}
\ee
which tend to $(\M m^2)^{\frac13} =\La$ as $n$ goes to $\infty$.\\

Interestingly, the same structure appears in static but non-spherically symmetric situations. To show this, let us consider again a configuration of the form \refeq{g}, but this time with $\n, \l$ and $\pi$ depending on $r$ and on the angle $\theta$. One then finds
\begin{displaymath}
(g^{-1} \tilde \eta)^{\mu}{}_{\nu }=\left(\begin{array}{c|c|c}
(-g_{tt})^{-1} \,& {\bf 0} \, &\, 0  \\
\hline
{\bf 0} & \,{\bf C} \, & \,{\bf 0} \\
\hline
0 \, &  {\bf 0} \, &\,d   \end{array}\right) 
\end{displaymath}
where 
\bea
C^r{}_r &=& g_{rr}^{-1} \left( (1-\tp'')^2   +\frac{1}{r^4}\left( r \tp'_{,\theta}-\tp_{,\theta}  \right)^2 \right) \\
C^r{}_\theta&=&  r^{-3} g_{rr}^{-1}  \left( r \tp'_{,\theta}-\tp_{,\theta}  \right)\left( r^2 \tp''+r \tp'+\tp_{,\theta \theta} -2 r^2  \right) \\
C^{\theta}{}_r &=&  r^{-5} \left(1+\frac{\pi}{\M}\right)^{-1}   \left( r \tp'_{,\theta}-\tp_{,\theta}  \right)\left( r^2 \tp''+r \tp'+\tp_{,\theta \theta} -2 r^2  \right)  \\
C^{\theta}{}_\theta &=&  \left(1+\frac{\pi}{\M}\right)^{-1} \left[    \left( 1-\frac{\tp'}{r}-\frac{\tp_{,\theta \theta}}{r^2}   \right)^2 +\frac{1}{r^4} \left(  r \tp'_{,\theta}-\tp_{,\theta} \right)^2 \right]
\eea
and
\be
d=\left(1+\frac{\pi}{\M}\right)^{-1}  \left(1-\frac{\tp'}{r} -{\rm Cot}(\theta) \frac{\tp_{,\theta}}{r^2}    \right)^2\,.
\ee
The discussion then proceeds completely similarly as above: when $r \tp'_{,\theta}-\tp_{,\theta} =0$, a matrix square-root of ${\bf C}$ is 
\be
{\rm Diag} \left( (g_{rr})^{-1/2} (1-\tp''), \left(1+\frac{\pi}{\M}\right)^{-1/2}   \left( 1-\frac{\tp'}{r}-\frac{\tp_{,\theta \theta}}{r^2}   \right) \right)\,,
\ee
$\gamma^{\mu}{}_{\nu}$ is linear in $\tp=\pi/\La^3$ and the minimal energy of interactions is $\M$. Additionally, when $\lambda=0$, an ``obvious'' square-root of ${\bf C}$ reads:
\begin{displaymath}
  \left(1+\frac{\pi}{\M}\right)^{-1/2}\left(\begin{array}{c|c}
1-\tp'' & -r^{-1}  (  r \tp'_{,\theta}-\tp_{,\theta}  ) \\
\hline
 -r^{-3}  (  r \tp'_{,\theta}-\tp_{,\theta}  )& 1-\frac{\tp'}{r}-\frac{\tp_{,\theta \theta}}{r^2}  \end{array}\right) 
\end{displaymath}
and the same conclusion holds. In the general case however, one can use Eq.~\refeq{trace-square-root} to find the trace of the relevant square root of ${\bf C}$, and hence an explicit expression for $\langle  \gamma \rangle$. Expanding the term $\M^2 m^2 \sqrt{-g} \langle \gamma \rangle$ in Eq.~\refeq{L-mass}, one then finds, in addition to the terms suppressed by $\M$ and $\Lambda_{1-2/(n-2)}$, an infinite number of interactions involving negative powers of $\La^3$, all proportional to $\l^2  \left( r \tp'_{,\theta}-\tp_{,\theta}  \right)^2$. They include for example interactions of the type
\be
b_n\, {\rm sin}(\theta)\,  \frac{ \l^2   \left( r \pi'_{,\theta}-\pi_{,\theta}  \right)^2}{r^2 \La^3 \M}  \frac{ \left(\pi''+\pi'/r+\pi_{,\theta \theta}/r^2  \right)^n }{\La^{3 n}}  \,,
\ee
with non-zero $b_n$ for all $n \geq 0$, whose energy scales are again given by Eq.~\refeq{E-n}. \\

As a summary, although there is no interaction at the energy $\La$ in the minimal model, and the lowest energy scale of interactions is $\M$ in static spherically symmetric configurations, interactions arising at energy scales arbitrarily close to $\La$ are present in generic spherically symmetric but time-dependent configurations, or generic static but non-spherically symmetric ones. In the non-minimal models of massive gravity, the existence of a \textit{finite number} of interactions arising at a \textit{well defined lowest energy scale} enables us to hope to capture the main physics of interactions by concentrating on this scale. This is the essence of the decoupling limit. Our analysis shows that no such hope is possible in the minimal model, either because there is no lowest energy scale of interaction --- this is the generic case --- or, in the presence of such a scale in SSS configurations, namely the Planck mass, because of the existence of an infinite number of interactions arising at this scale, including for instance the ones of General Relativity.

\section{Exact equations of motion and obstruction to a Vainshtein mechanism}
\label{Exact}

In the previous section, we have shown that in static spherically symmetric configurations, interactions below the Planck mass between $\bar h\mn$ and $\pi$ vanish identically. This remarkable fact points towards the absence of a Vainshtein mechanism in this set up, but does not prove it. In this section, we settle this question by resorting to the exact equations of motion. Considering generic vacuum stationary and spherically symmetric configurations, we show, within these hypotheses, that there is an obstruction that prevents the recovery of General Relativity in the minimal model.

By definition, a stationary and spherically symmetric configuration with Minkowski reference metric reads, in Lorentzian coordinates:
\bea
g\mn d x^\mu d x^\nu&=&-a^2(r) dt^2+2 d(r) dt dr +b^2(r) dr^2+c^2(r) d\Omega^2 \\
\eta\mn d x^\mu d x^\nu&=&-dt^2+dr^2+r^2 d\Omega^2\,.
\eea
To enable the comparison with the Schwarzschild solution of General Relativity, we define the four functions $\lambda(r), \mu(r), \nu(r), \alpha(r)$ out of the four functions $a(r), b(r), c(r), d(r)$ as follows:
\bea
a^2(r)&=&e^{\nu(r)}\,,\quad c^2(r)=r^2 e^{\mu(r)}\,,\quad b^2(r)+\frac{d^2(r)}{a^2(r)}=e^{\lambda(r)+\mu(r)} \left(1+\frac{r}{2} \frac{d \mu}{d r} \right)^2 \\
\frac{d(r)}{a^2(r)}&=& \alpha(r) e^{\frac{\mu(r)}{2}} \left(1+\frac{r}{2} \frac{d \mu}{d r} \right)\,.
\eea
We then make the change of coordinates $(t,r) \to (T,R)$ such that
\bea
R &=&c(r)=r e^{\frac{\mu(r)}{2} } \label{R-r}\\
d T& =& dt- \frac{d(r)}{a^2(r)} dr\,,
\eea
giving
\bea
g_{\mu \nu} dx^{\mu} dx^{\nu} &=& -e^{\nu(R)} dT^2+e^{\lambda(R)} dR^2+R^2 d \Omega^2 \\
\eta_{\mu \nu} dx^{\mu} dx^{\nu} &=& -(dT+\alpha(R) dR )^2+\left(1-\frac{R \mu'(R)}{2} \right)^2 e^{-\mu(R)} dR^2+ e^{-\mu(R)}R^2 d \Omega^2
\eea
(where $\lambda(R)=\lambda(r(R))$ and similarly for $\mu, \nu, \alpha$). This is the most general form compatible with spherical symmetry and stationarity given in Ref.~\cite{Damour:2002gp}, of which we follow the notations. It is particularly convenient to deal with as it separates the directly observable gravitational variables $\nu(R), \lambda(R)$ from the ``gauge'' functions $\mu(R), \alpha(R)$, which enter only the unobservable reference metric. In this gauge, the Schwarzschild solution of General Relativity reads 
\be
\nu_{{\rm GR}}=-\lambda_{{\rm GR}}= {\rm ln} \left(1-\frac{R_S}{R} \right)\,,
\label{GR}
\ee
where $R_S$ is the Schwarzschild radius of the source.

Like in section \ref{Energy}, one can determine an explicit expression for the relevant square root of $(g^{-1} f)^{\mu}{}_{\nu}$:
\begin{displaymath}
\gamma^{\mu}{}_{\nu }=\left(\begin{array}{c|c}
{\bf D}\, & {\bf 0}  \\
\hline
{\bf 0} &\, {\bf E} \end{array}\right) \,,
\end{displaymath}
where
\begin{displaymath}
{\bf D}=\frac{1}{t}\left(\begin{array}{c|c}
e^{-\nu}+s & e^{-\nu} \alpha  \\
\hline
- e^{-\lambda} \alpha &  e^{-\lambda} \left( e^{-\mu}(1- \frac{R\mu'}{2})^2   -\alpha^2\right) +s  \end{array}\right) \,,
\end{displaymath}
\be
{\bf E}=e^{-\mu/2}  \bf{1}\,,
\ee
with
\bea
s&=&\frac12 e^{-(\lambda+\mu+\nu)/2}(-2+R \mu') \\
t&=&\left[\frac14 e^{-(\lambda+\mu+\nu)} \left( 2e^{(\lambda+\mu)/2}-2 e^{\nu/2}+e^{\nu/2} R \mu' \right)^2-e^{-\lambda} \alpha^2 \right]^{1/2}\,.
\eea
As $\alpha$ only enters the action through the mass term in Eq.~\refeq{L-beta}, and moreover non-derivatively, its equation of motion is purely algebraic. It reads, for any model of dRGT massive gravity:
\be
\alpha \left(\beta_1 e^{\mu}+4 \beta_2 e^{\mu/2}+6 \beta_3  \right)=0\,.
\label{alpha-eom}
\ee
 Although $\alpha=0$ is always a solution, there may exist other solutions for the non-minimal models with $\beta_2$ or/and $\beta_3$ non-zero, depending on the signs and relative values of the parameters. However, in the minimal model --- $\beta_1=-2, \beta_2=\beta_3=0$ --- Eq.~\refeq{alpha-eom} readily gives $\alpha=0$. In other words, all spherically symmetric and stationary solutions of the minimal model are actually static.

To proceed further with the remaining variables $\lambda, \mu, \nu$, let us write down the Einstein equations of motion in vacuum: 
\be
G\mn=m^2 T\mn^{{\rm mass}}\,,
\label{eoms}
\ee
where the energy-momentum tensor derived from the mass term in Eq.~\refeq{L-mass} reads \cite{Volkov:2013roa}
\be
T\mn^{{\rm mass}}=3 g\mn+\gamma\mn-\langle \gamma \rangle g\mn\,.
\ee
The $(T,T)$ and $(R,R)$ components of Eq.~\refeq{eoms} read respectively:
\bea
e^{\nu-\lambda} \left(\frac{\lambda'}{R}+\frac{1}{R^2}(e^{\lambda}-1)  \right) &=& m^2 T_{T T}^{{\rm mass}} 
\,, \label{TT}\\
\frac{\nu'}{R}+\frac{1}{R^2}(1- e^{\lambda}) &=& m^2 T_{R R}^{{\rm mass}}
\label{RR} \,,
\eea
where
\bea
T_{T T}^{{\rm mass}}  &=&-\frac12 e^{\nu-\frac12 (\lambda+\mu)} \left(-2-4\, e^{\lambda/2}+6 \,e^{\frac12 (\lambda+\mu)}+R\, \mu' \right) \,, \\
 T_{R R}^{{\rm mass}}&=& -e^{\lambda}(-3+2\, e^{-\mu/2}+e^{-\nu/2}) \,.
\eea
To supplement these equations, one can use the $(\theta,\theta)=(\phi,\phi)$ equation of motion. Alternatively, when Eqs.~\refeq{TT}-\refeq{RR} are satisfied, it is equivalent to the Bianchi identity
\be
f_g \equiv \nabla^{\mu} T_{\mu R}^{{\rm mass}}=0\,,
\label{Bianchi}
\ee
where
\be
f_g= \left(2 R\, e^{\frac12 (\lambda+\mu)} \right)^{-1} \left(1-\frac{R \, \mu'}{2}\right) \left( 4-4 e^{\lambda/2}+R\, \nu' \right)\,.
\ee
It is excluded that $1-R \, \mu'(R)/2$ vanishes, as it would correspond to a non-invertible change of coordinate \refeq{R-r}. The equation \refeq{Bianchi} therefore implies that $4-4 e^{\lambda/2}+R \,\nu'=0$. Combining this relation with Eq.~\refeq{RR}, one obtains the algebraic relation
\be
4 e^{-\lambda/2}-1-3 e^{-\lambda}=-m^2 R^2  (-3+2 e^{-\mu/2}+e^{-\nu/2}) \,,
\ee
from which one obtains $\mu$ in terms of $\nu$ and $\lambda$. Inserting this into Eq.~\refeq{TT}, one finally obtains the coupled system of two first-order differential equations for $\lambda(R)$ and $\nu(R)$:
\bea
R\, \lambda'- e^{\lambda} (m^2 R^2-3)+m^2 R^2 \, e^{\lambda-\nu/2 }+1- 4\, e^{\lambda/2}&=&0\,, \label{Eq1} \\
R \,\nu'+4-4 \,e^{\lambda/2}&=&0 \label{obstruction} \,.
\eea
Linearizing these equations, one can obtain the solutions
\bea
\lambda_{L}&=&\frac{2 C R_S}{3R}(1+m R) e^{-m R}\,, \label{lambda-L}\\
\nu_{L}&=&-\frac{4 C R_S}{3R} e^{-m R}  \label{nu-L}   \,,
\eea
where $C$ is a constant of integration. At large radii, they display the standard $e^{-m R}$ Yukawa-suppression expected from a massive graviton. At small radii, $R \ll m^{-1}$, one gets $\nu_L \sim -2 \lambda_L$, to be contrasted with the Schwarzschild result $\nu_{{\rm GR}}=-\lambda_{{\rm GR}}$ \refeq{GR}. This is a manifestation of the famous vDVZ (van Dam-Veltman-Zakharov) discontinuity \cite{vDVZ}, namely that the massless limit of Fierz-Pauli massive gravity does not coincide with General Relativity. In this respect, the crucial observation made by Vainshtein \cite{Vainshtein:1972sx} is that the linear approximation breaks down for $R \lesssim R_V$, where the Vainshtein radius $R_V$ grows to infinity as the mass $m$ approaches $0$\footnote{The expression of the Vainshtein radius depends on the theory of massive gravity under consideration \cite{Babichev:2009us}. It reads $R_V=\left( R_S/m^4\right)^{1/5}$ in generic massive gravity theories and $R_V=\left( R_S/m^2\right)^{1/3}$ in dRGT massive gravity.}. As a result, the linear approximation of massive gravity is nowhere applicable in the massless limit, giving hope that the vDVZ discontinuity is merely an artifact of the linear perturbation theory, and that the full non-linear solution displays a smooth limit with General Relativity. In particular, for a theory of massive gravity to be observationally relevant, its solution should be very close to the one of GR inside the solar system, where the former has been tested with very fine accuracy. Following this idea, Vainshtein suggested to look for SSS solutions of massive gravity, at least sufficiently close to the source, as an expansion in powers of the graviton mass around the Schwarzschild solution:
\be
X(R)=\sum_{n=0}^{\infty} m^{2n} X_n(R)\,, \quad {\rm with} \quad \lambda_0=\lambda_{{\rm GR}} \quad {\rm and} \quad \nu_0=\nu_{{\rm GR}}\,,
\label{expansion-m2}
\ee
where $X$ collectively stands for $\lambda, \mu, \nu$ and the $X_n$ do not depend on $m$. That such solutions exist for $R \ll R_V$, \textit{and} can be extended globally to match the solutions of the linearized theory \refeq{lambda-L}-\refeq{nu-L} for $R \gg R_V$, has been shown for the first time for some massive gravity models in Refs.~\cite{Babichev:2009us,Babichev:2009jt,Babichev:2010jd}, establishing that the Vainshtein mechanism can work in the context of spherically symmetric solutions (see the earlier work \cite{Deffayet:2001uk} in the context of cosmology).

However, in the case of the minimal model of interest here, there is already an obstruction to find solutions of the form \refeq{expansion-m2} at the zeroth-order. The massless limits of the Einstein equations \refeq{TT}-\refeq{RR} are obviously satisfied by $\nu_{{\rm GR}}$ and $\lambda_{{\rm GR}}$ by definition. However, the latter should also verify the extra equation \refeq{Bianchi}, which does not apply in General Relativity. In generic models of massive gravity, this is not problematic: plugging them into this equation, one obtains a differential equation that $\mu_0(R)$ has to satisfy. However, the peculiarity of the minimal model is that the Bianchi identity \refeq{Bianchi} leads to an equation that does not involve $\mu$, but $\nu$ and $\lambda$ only, namely Eq.~\refeq{obstruction}. As $\nu_{{\rm GR}}$ and $\lambda_{{\rm GR}}$ \refeq{GR} do not verify this equation, an exact solution of the system \refeq{Eq1}-\refeq{obstruction} cannot reduce to the Schwarzschild solution in the massless limit. Therefore, we conclude that the Vainshtein mechanism is ineffective in stationary and spherically symmetric configurations of the minimal model.

\section{Discussion} 
\label{Conclusion}

If one wishes that modified theories of gravity have somewhat substantial effects on cosmological scales, and that they reproduce the successful phenomenology of General Relativity in the solar system, they must come with screening mechanisms that enable to hide the effects of their additional degrees of freedom compared to GR on solar system/laboratory scales. In massive gravity, the Vainshtein mechanism plays this role. In this paper, we have studied it in the so-called minimal model of dRGT massive gravity. In particular, we have shown in section \ref{Exact} that its non-linearities are inefficient to restore the continuity with General Relativity in stationary and spherically symmetric configurations, in other words that the Vainshtein mechanism is ineffective under these hypotheses. To reach this conclusion, we did not need to find explicitly the corresponding vacuum solutions, although it could be interesting to determine them, exactly or numerically.\\

Probably more important are the consequences of our study of the energy scales of interactions in section \ref{Energy}. The minimal model being characterised by the absence of interactions in the decoupling limit, \textit{i.e.} at the lowest possible energy $\La$, we investigated their structure at higher energies, concentrating on the interactions between the helicity-2 modes and the scalar graviton. In this framework, we proved the remarkable fact that in static spherically symmetric configurations, the lowest energy scale of interactions is pushed up to the Planck mass. However, we have also shown the peculiarity of these configurations: in generic non spherically symmetric or time-dependent situations, interactions at energies arbitrarily close to $\La$ reappear. Although it is hard to reach conclusions solely on these facts, one can thus wonder whether the small degree of spherical symmetry breaking in the solar system can be enough to lead to a successful Vainshtein mechanism in the minimal model. More generally, while screening mechanisms have been mostly studied in static/stationary spherically symmetric situations up to now (see however Ref.~\cite{Babichev:2011iz}), our analysis leads us to question whether the high degree of symmetry of these configurations might miss some important physical phenomena that arise in nature in realistic circumstances. These interesting questions are left for future research.

\subsection*{Acknowledgements} 

I would like to thank Claudia de Rham for initial collaboration and for insightful comments on a draft version of this paper, and C\'edric Deffayet, David Langlois and Lorenzo Sorbo for discussions related to the topic of this paper. This work was supported by French state funds managed by the ANR
within the Investissements d'Avenir programme under reference
ANR-11-IDEX-0004-02.


\begin{thebibliography}{99}



\bibitem{Weinberg:1988cp}
  S.~Weinberg,
  Rev.\ Mod.\ Phys.\  {\bf 61} (1989) 1.



\bibitem{Dvali:2002pe}
  G.~Dvali, G.~Gabadadze and M.~Shifman,
  Phys.\ Rev.\ D {\bf 67} (2003) 044020
  [hep-th/0202174].

\bibitem{ArkaniHamed:2002fu}
  N.~Arkani-Hamed, S.~Dimopoulos, G.~Dvali and G.~Gabadadze,
  hep-th/0209227.

\bibitem{Dvali:2007kt}
  G.~Dvali, S.~Hofmann and J.~Khoury,
  Phys.\ Rev.\ D {\bf 76} (2007) 084006
  [hep-th/0703027 [HEP-TH]].



 \bibitem{Boulware:1973my}
  D.~G.~Boulware and S.~Deser,
  Phys.\ Rev.\  D {\bf 6}, 3368 (1972).


 \bibitem{Creminelli:2005qk}
  P.~Creminelli, A.~Nicolis, M.~Papucci and E.~Trincherini,
  JHEP {\bf 0509}, 003 (2005).
  [arXiv:hep-th/0505147].

\bibitem{Deffayet:2005ys}
  C.~Deffayet and J.~-W.~Rombouts,
  Phys.\ Rev.\ D {\bf 72} (2005) 044003
  [gr-qc/0505134].



\bibitem{deRham:2010ik}
  C.~de Rham, G.~Gabadadze,
  Phys.\ Rev.\  {\bf D82}, 044020 (2010).
  [arXiv:1007.0443 [hep-th]].


\bibitem{deRham:2010kj}
  C.~de Rham, G.~Gabadadze, A.~J.~Tolley,
  Phys.\  Rev.\  Lett.\  106, {\bf 231101} (2011).
  [arXiv:1011.1232 [hep-th]].






\bibitem{Hassan:2011hr}
  S.~F.~Hassan, R.~A.~Rosen,
  [arXiv:1106.3344 [hep-th]].


\bibitem{deRham:2011rn}
  C.~de Rham, G.~Gabadadze and A.~J.~Tolley,
  Phys.\ Lett.\ B {\bf 711}, 190 (2012)  [arXiv:1107.3820 [hep-th]].  


\bibitem{deRham:2011qq}
  C.~de Rham, G.~Gabadadze and A.~J.~Tolley,
  JHEP {\bf 1111} (2011) 093  [arXiv:1108.4521 [hep-th]].  



\bibitem{Hassan:2011ea}
  S.~F.~Hassan and R.~A.~Rosen,
  arXiv:1111.2070 [hep-th].  



\bibitem{Mirbabayi:2011aa}
  M.~Mirbabayi,
  arXiv:1112.1435 [hep-th].  


\bibitem{Golovnev:2011aa}
  A.~Golovnev,
  Phys.\ Lett.\ B {\bf 707} (2012) 404
  [arXiv:1112.2134 [gr-qc]].


\bibitem{Hassan:2012qv}
  S.~F.~Hassan, A.~Schmidt-May and M.~von Strauss,
  Phys.\ Lett.\ B {\bf 715} (2012) 335
  [arXiv:1203.5283 [hep-th]].

\bibitem{Kluson:2012wf}
  J.~Kluson,
  Phys.\ Rev.\ D {\bf 86} (2012) 044024
  [arXiv:1204.2957 [hep-th]].
  
\bibitem{Deffayet:2012zc}
  C.~Deffayet, J.~Mourad and G.~Zahariade,
  JHEP {\bf 1303} (2013) 086
  [arXiv:1208.4493 [gr-qc]].



\bibitem{Will:2005va}
  C.~M.~Will,
  Living Rev.\ Rel.\  {\bf 9}, 3 (2005)  [gr-qc/0510072].  




\bibitem{vDVZ}
H.~van Dam and M.~J.~G.~Veltman,
  Nucl.\ Phys.\  B {\bf 22}, 397 (1970);
V.~I.~Zakharov,
  JETP Lett.\  {\bf 12} (1970) 312
  [Pisma Zh.\ Eksp.\ Teor.\ Fiz.\  {\bf 12} (1970) 447].


 \bibitem{Vainshtein:1972sx}
  A.~I.~Vainshtein,
  Phys.\ Lett.\  B {\bf 39}, 393 (1972);





\bibitem{Babichev:2009us}
  E.~Babichev, C.~Deffayet and R.~Ziour,
  JHEP {\bf 0905} (2009) 098
  [arXiv:0901.0393 [hep-th]].

\bibitem{Babichev:2009jt}
  E.~Babichev, C.~Deffayet and R.~Ziour,
  Phys.\ Rev.\ Lett.\  {\bf 103} (2009) 201102
  [arXiv:0907.4103 [gr-qc]].

\bibitem{Babichev:2010jd}
  E.~Babichev, C.~Deffayet and R.~Ziour,
  Phys.\ Rev.\ D {\bf 82} (2010) 104008
  [arXiv:1007.4506 [gr-qc]].

\bibitem{Babichev:2013usa}
  E.~Babichev and Céd.~Deffayet,
  Class.\ Quant.\ Grav.\  {\bf 30} (2013) 184001
  [arXiv:1304.7240 [gr-qc]].



\bibitem{Koyama:2011xz}
  K.~Koyama, G.~Niz and G.~Tasinato,
  Phys.\ Rev.\ Lett.\  {\bf 107} (2011) 131101
  [arXiv:1103.4708 [hep-th]].




\bibitem{Nieuwenhuizen:2011sq}
  T.~.M.~Nieuwenhuizen,
  Phys.\ Rev.\ D {\bf 84} (2011) 024038
  [arXiv:1103.5912 [gr-qc]].


\bibitem{Koyama:2011yg}
  K.~Koyama, G.~Niz and G.~Tasinato,
  Phys.\ Rev.\ D {\bf 84} (2011) 064033
  [arXiv:1104.2143 [hep-th]].



\bibitem{Chkareuli:2011te}
  G.~Chkareuli and D.~Pirtskhalava,
  Phys.\ Lett.\ B {\bf 713} (2012) 99
  [arXiv:1105.1783 [hep-th]].



\bibitem{Gruzinov:2011mm}
  A.~Gruzinov and M.~Mirbabayi,
  Phys.\ Rev.\ D {\bf 84} (2011) 124019
  [arXiv:1106.2551 [hep-th]].

\bibitem{Comelli:2011wq}
  D.~Comelli, M.~Crisostomi, F.~Nesti and L.~Pilo,
  Phys.\ Rev.\ D {\bf 85} (2012) 024044
  [arXiv:1110.4967 [hep-th]].

\bibitem{Berezhiani:2011mt}
  L.~Berezhiani, G.~Chkareuli, C.~de Rham, G.~Gabadadze and A.~J.~Tolley,
  Phys.\ Rev.\ D {\bf 85} (2012) 044024
  [arXiv:1111.3613 [hep-th]].



\bibitem{Sjors:2011iv}
  S.~Sjors and E.~Mortsell,
  arXiv:1111.5961 [gr-qc].




\bibitem{Sbisa:2012zk}
  F.~Sbisa, G.~Niz, K.~Koyama and G.~Tasinato,
  Phys.\ Rev.\ D {\bf 86} (2012) 024033
  [arXiv:1204.1193 [hep-th]].



\bibitem{Volkov:2013roa}
  M.~S.~Volkov,
  Class.\ Quant.\ Grav.\  {\bf 30} (2013) 184009
  [arXiv:1304.0238 [hep-th]].



\bibitem{ArkaniHamed:2002sp}
N.~Arkani-Hamed, H.~Georgi and M.~D.~Schwartz,
Annals Phys.\  {\bf 305}, 96 (2003).



\bibitem{Gratia:2013gka}
  P.~Gratia, W.~Hu and M.~Wyman,
  Class.\ Quant.\ Grav.\  {\bf 30} (2013) 184007
  [arXiv:1305.2916 [hep-th]].





 \bibitem{Fierz:1939ix}
  M.~Fierz and  W.~Pauli,
  Proc.\ Roy.\ Soc.\ Lond.\  A {\bf 173}, 211 (1939).




\bibitem{Siegel:1993sk}
  W.~Siegel,
  ``Hidden gravity in open string field theory,''
  Phys.\ Rev.\  {\bf D49}, 4144-4153 (1994).
  [hep-th/9312117].


\bibitem{Berezhiani:2013dw}
  L.~Berezhiani, G.~Chkareuli and G.~Gabadadze,
  Phys.\ Rev.\ D {\bf 88} (2013) 124020
  [arXiv:1302.0549 [hep-th]].


\bibitem{Berezhiani:2013dca}
  L.~Berezhiani, G.~Chkareuli, C.~de Rham, G.~Gabadadze and A.~J.~Tolley,
  Class.\ Quant.\ Grav.\  {\bf 30} (2013) 184003
  [arXiv:1305.0271 [hep-th]].





\bibitem{Damour:2002gp}
  T.~Damour, I.~I.~Kogan and A.~Papazoglou,
  Phys.\ Rev.\ D {\bf 67} (2003) 064009
  [hep-th/0212155].



\bibitem{Deffayet:2001uk}
  C.~Deffayet, G.~R.~Dvali, G.~Gabadadze and A.~I.~Vainshtein,
  Phys.\ Rev.\ D {\bf 65} (2002) 044026
  [hep-th/0106001].



\bibitem{Babichev:2011iz}
  E.~Babichev, C.~Deffayet and G.~Esposito-Farese,
  Phys.\ Rev.\ Lett.\  {\bf 107} (2011) 251102
  [arXiv:1107.1569 [gr-qc]].

  


\end{thebibliography}
\end{document}